\documentclass[11pt,a4paper]{article}
\begin{document}
\title{ The effect of the tortoise coordinates  on the tunnel effect
\footnote{ E-mail of Tian:
 hua2007@126.com, tgh-2000@263.net}}
\author{Tian Gui-hua$^{a,b}$ ,\ \   Zhao Zheng$^{c}$,\ \ Shi-kun Wang$^{b}$\\
a. School of Science, Beijing University \\
of Posts And Telecommunications, Beijing 100876, China. \\ b.
Academy of Mathematics and Systems Science,\\ Chinese Academy of
Sciences,(CAS) Beijing 100080, China.\\
c. Department of Physics, Beijing Normal University, Beijing
100875, China}
\date{}
\maketitle

\begin{abstract}
The tunnel process of the quantum wave from the light cone is
carefully discussed. They are applied in the massive quantum
particles from the Schwarzschild black hole in the Kruskal metric.
The tortoise coordinates prevent one from understanding the tunnel
process, and are investigated with care. Furthermore, the massive
particles could come out of the black hole either by the Hawking
radiation or by the tunnel effect; the tunnel effect might give
more information about the black hole.


\end{abstract}

  When the potential $V$ in some region is greater than the energy
$E$, the classical particles could not pass through the potential
barrier and obtain admission into the other region with potential
lower than the energy. But in quantum mechanics, the classical
forbidden region could be penetrated and passed
through\cite{land}. This case belongs to the non-relativistic
quantum theory. Now, consider the relativistic case, the classical
particle must move inside its light cone. The classical forbidden
region for the particle is the outside of its future light cone.
Similarly, the relativistic quantum wave equations are the K-G
equations, etc. When the initial mass $\mu$ is not zero for the
particle, the particle's phase velocity is greater than the
light's velocity. The wave can not be confined inside of the
future light cone, the quantum particle could pass through the
light cone. We show the tunnel process by an example of the K-G
particle in two dimensional Minkowski space time
\begin{equation}
ds^{2}=-dT^{2}+dZ^{2}.\label{2-dim minkowski}
\end{equation}
The massive K-G equation is
\begin{equation}
-\frac{\partial^{2}\Psi}{\partial T^{2}}+ \frac {\partial ^2
\Psi}{\partial Z^2}-\mu^{2}\Psi=0.\label{kg in2-dim  m}
\end{equation}
The mode-decomposition solutions are
\begin{equation} \Psi
=e^{-i\omega T+ikZ}\label{mode in 2-dim m}
\end{equation}
with the relation The mode-decomposition solutions are
\begin{equation}
\omega ^2=k^2+\mu^2.\label{mode relation in 2-dim m}
\end{equation}
From the Eq.(\ref{mode relation in 2-dim m}), we see the phase
velocity $\frac{\omega}{k}>1$ whenever the initial mass $\mu \ne
0$.

The effect could be used in a wider range in physics. We first
consider its application in the Minkowski space time. The 4-dim
Minkowski space time is defined by
\begin{equation}
ds^{2}=-dT^{2}+dZ^{2}+dx^{2}+ dy^{2},\label{orimetric-m}
\end{equation}
and consists of four parts. Part $I$, part $II$, part $I'$, part
$II'$ correspond to $T^2-Z^2<0,\ Z>0$; $T^2-Z^2>0,\ T>0$;
$T^2-Z^2<0,\ Z<0$; $T^2-Z^2>0,\ T<0$ respectively.

The Rindler metric is
\begin{equation}
ds^{2}=-(1+az)dt^{2}+(1+az)^{-1}dz^{2}+dx^{2}+
dy^{2}.\label{orimetric r}
\end{equation}
The constant spatial coordinates with $z<-\frac 1a$  are also the
world lines of constantly accelerated observers. The null
geodesics have constant $z$ axis component $z=-\frac 1a$ with its
acceleration going to infinity. These null geodesics consist of
the horizon of the Rindler space time.

The Rindler space time corresponds to the part $I$ in the
Minkowski space time, and the part $I'$ is not communicable with
it. The parts $II$ and $II'$ are the future and past of the
Rindler space time respectively.

Because any classical particle can not surpass the light,
classical particles in the Part $II$ can not enter the Rindler
space time later. Nevertheless, applying the above tunnel process,
we get the massive particle could enter the Rindler space time
from the part $II$. The effect is only for massive K-G particle,
and is different from the Unruh effect.  The massless K-G
particles have not such effect of tunnel, but they have Unruh
effect.

Because the Rindler space time is similar in many respects with
the Schwarzschild black hole, the effect of the tunnel could also
extend to the Schwarzschild black hole.

Why researchers have not noticed the effect of the tunnel for the
massive K-G particles in the Rindler space time? How to extend it
to the Schwarzschild black hole? The paper mainly address the
problems.

The tortoise coordinates of the Rindler space time and the
Schwarzschild black hole  prevent one from finding the tunnel
process from the black hole or part $II$  for the massive
particles.

In fact, the researcher found the Hawking Radiation or the Unruh
effect for over thirty years,  why have not they found the tunnel
effect for the massive K-G particles yet? The reasons are the
following.

Generally, even the quantum effects are considered, it is taken
for granted that the physical entities in part $II$ have no
influence on the Rindler space time. The thought is regarded right
until the appearance of the Hawking Radiation. This is mainly due
to the form of the K-G equation in the Rindler space time
containing the tortoise coordinate $t$.

The K-G equation in the Rindler metric (\ref{orimetric r}) reads
\begin{equation}
-\frac1{1+az}\frac{\partial^2\psi}{\partial t^2}+ \frac {\partial
}{\partial z}\left[(1+az)\frac {\partial \psi }{\partial
z}\right]- (k_1^2+k_2^2+\mu^2)\psi=0.\label{kg in r2}
\end{equation}
Define the spatial tortoise coordinate $z_*$ as
\begin{equation}
z_*=\frac 1a\ln(1+az)
\end{equation}
which makes the horizon $1+az=0$  and the spatial infinity into
$z_*\rightarrow -\infty$ and $z_*\rightarrow +\infty$
respectively. The Eq.(\ref{kg in r2}) then becomes
\begin{equation}
\frac{\partial^2\psi}{\partial t^2}- \frac {\partial ^2
\psi}{\partial z_*^2}+(1+az) (k_1^2+k_2^2+\mu^2)\psi =0.\label{kg
in r3}
\end{equation}
From the Eq.(\ref{kg in r3}), we see that the velocity of the
scalar particle cannot exceed speed limit at the horizon $z=-\frac
1a $. It is the same as the light's velocity. Therefore, it is
natural to regard the Rindler space time as independent entity.
This is wrong and discussed carefully in Ref\cite{tian6R}.

In fact, the Eq.(\ref{kg in r3}) gives less information at the
horizon, and it can mislead one to error. From it, we obtain the
wrong information that even the massive Klein-Gordon particle
could not tunnel from the horizon. Whereas the tunnel process is
very simple in the Minkowski coordinates. It is the Eq.(\ref{kg in
r3}) that make the tunnel process impossible. The Eq.(\ref{kg in
r3}) is connected with the Rindler metric, or the accelerated
frame. In the Rindler metric, the time coordinate $t$ is a
tortoise coordinate, and it could result in many difficult
explanation of the physical process. Actually, the future and past
horizons correspond to $t\rightarrow +\infty $, $t\rightarrow
-\infty $ respectively. This is just the manifest of the tortoise
property of the Rindler time coordinate $t$. So, the process
involved the horizon is difficult to describe by the Rindler
coordinates.

We reinforce our viewpoint again: the physical process can be
easily explained by the Minkowski metric, it is only the Rindler
tortoise coordinate $t$ that make the explanation of the physical
process difficult.

The Unruh effect corresponds to the excitation of the modes whose
phase velocity is equal to that of the light at the
horizon\cite{unru}. It is really curious that the excited modes
are not those  whose  phase velocity surpass the light.

Furthermore, the phase velocity could be greater than the light
velocity even in the Rindler space time whenever the scalar field
is not at the horizon.

The same situations exist for the Schwarzschild black hole. We now
extend the tunnel effect to it.

The Schwarzschild black hole is described by the metric
\begin{equation}
ds^{2}=-(1-\frac{2m}{r})dt^{2}+(1-\frac{2m}{r})^{-1}dr^{2}+r^{2} d
\Omega ^{2}.\label{orimetric}
\end{equation}
The observers $r=r_{0}, \theta =\theta_{0}, \varphi=\varphi_{0}$
have constant acceleration
\begin{equation}
A=\frac{m}{r^{2}}(1-\frac{2m}{r})^{-\frac{1}{2}}. \label{a}
\end{equation}
When $r_{0}\rightarrow 2m$, the time-like curve $r=r_{0}, \theta
=\theta_{0}, \varphi=\varphi_{0}$ becomes the null geodesic with
its proper acceleration $A\rightarrow \infty$. The null geodesics
$r_{0}=2m, \theta =\theta_{0}, \varphi=\varphi_{0}$with $0 \leq
\theta \leq \pi, 0 \leq \varphi \leq 2\pi $ consist the horizon of
these observers $r=r_{0} <2m, \theta =\theta_{0},
\varphi=\varphi_{0}$. Of course, it is also the horizon of the
black hole.

Nevertheless, it is obvious that the Schwarzschild metric as a
whole is the same with that of the Rindler metric.

The Schwarzschild metric and the Rindler metric all are in the
accelerated frame and have the same kind horizons for those
accelerated observers.

Furthermore, the Schwarzschild metric has the complete space time,
that is, the Kruskal space time. The Rindler's completion is the
Minkowski space time. The geometric properties of the
Schwarzschild black hole are the same as those of the Rindler
space time.

These striking similarities stimulate researcher to investigate
their further connection.

Later, Hawking showed that the black hole could radiate thermal
radiation with temperature, therefore he made great progress of
black hole study. Unruh investigated the radiation connected with
the Rindler space time, and obtained that the Rindler observers
could also detect the thermal radiation with the temperature k
equal to its surface gravity too. Unruh appeared first to notice
the same origin of the Hawking Radiation and the Unruh effect.
Actually, the two effects all originate from the horizon as the
accelerated observers.

We now study the tunnel effect in the Schwarzschild black hole.
 The K-G equations in the Schwarzschild metric (\ref{orimetric})
is
\begin{equation}
\frac{\partial^{2}Q}{\partial t^2}-\frac{\partial^{2}Q}{\partial
r^{*2}}+\left(1-\frac{2m}{r}\right)\left[\frac{l\left(l+1\right)}{r^{2}}
+\frac{2m}{r^{3}}+\mu^2\right]Q=0\label{rwscalar}
\end{equation}
where $\mu$ is its mass. The third term in the Eq.(\ref{rwscalar})
is zero at the horizon. This equation (\ref{rwscalar}) has a
striking characteristic: its phase velocity equals to that of
light at the horizon. This has influence not only on the stable
study, but also for the explanation of the tunnel
effect\cite{tiantortoise}. Therefore, one is easily led to obtain
the conclusion that the scalar field inside of the black hole
($r<2m$) could not penetrate through
 the horizon and enter into the region $r>2m$. This fact is
truly wrong caused by the tortoise coordinates $t,\ r_*$.
Rewriting the scalar perturbation equation in the Kruskal
coordinates
\begin{equation}
ds^{2}=\frac{32m^3}{r}e^{-\frac
r{2m}}\left[-dT^{2}+dX^{2}\right]+r^{2} d \Omega
^{2},\label{kruskalmetric}
\end{equation}
and by the decomposition of the variables $Q=\psi (T,X)Y_{lm'}$
where $Y_{lm'}$ are the spherical harmonic functions, we obtain
\begin{equation}
\frac{\partial^{2}\psi}{\partial
T^2}-\frac{\partial^{2}\psi}{\partial
X^{2}}+\left(\frac{32m^3}{r}e^{-\frac
r{2m}}\right)\left[-\frac{T}{2mr}\frac{\partial \psi }{\partial T
}-\frac{X}{2mr}\frac{\partial \psi }{\partial X
}+\mu^2+\frac{l\left(l+1\right)}{r^{2}}\right]\psi=0.\label{scalar
in kruskal}
\end{equation}
By the geometric approximation at the horizon, we obtain the
asymptotic form for out-going wave $\psi$
\begin{equation}
\psi =A'e^{i\omega T-ikX}\label{psi-2m}
\end{equation}
with the relation
\begin{equation}
\omega
^2=k^2+16m^2e^{-1}\left[\mu^2+\frac{l\left(l+1\right)}{4m^2}\right]\label{disper
relation }
\end{equation}
under the condition $|\omega |\gg 4e^{-1}T$ at the time $T$.

Therefore, the phase velocity at the horizon
\begin{equation}
V_p=\pm \frac{\omega }{k}\label{V-2m}
\end{equation}
is greater than $c=1$ of the light whenever the mass $\mu$ or $l$
is not equal to zero. This is just the tunnel effect.

The excited modes in Hawking radiation are those whose phase
velocity are the light velocity. The tunnel effect is different
from the Hawking radiation.

\section*{Some comments}

It is shown that the tunnel process exists for the massive K-G
particles, it can also  easily extend to massive particles of
non-zero spin. This may have some effect on the black hole's
formation. The massive particles could come out of the black hole
either by the Hawking radiation or by the tunnel effect, it might
give some information about the black hole.

\section*{Acknowledgments}

Supported by the national Natural Science Foundations of China
under Grant  No.10475013, No.10373003, No.10375087; the National
Key Basic Research and Development Programme of China under Grant
No. 2004CB318000
 and the post-doctor foundation of
China.

\end{document}